\documentclass[preprint,showpacs,prd,floatfix,nofootinbib]{revtex4-1}
\usepackage{graphicx}
\usepackage{amssymb}
\usepackage{amsmath}
\usepackage{bm}
\usepackage{natbib}
\newcommand{\p}[1]{(\ref{#1})}
\def\Journal#1#2#3#4{{#1} {\bf #2}, #3 (#4)}

\def\npa{{Nucl. Phys.} A}

\def\plb{{Phys. Lett.}  B}
\def\prl{Phys. Rev. Lett.}
\def\prc{{Phys. Rev.} C}
\def\prd{{Phys. Rev.} D}
\def\aap{Astron. Astrophys.}
\def\apj{Astrophys. J.}

\def\ptp{Prog. Theor. Phys.}
\def\ppnp{Prog. Part. Nucl. Phys.}
\def\jpg{{J. Phys.} G}
\def\jpc{{J. Phys.} C}
\def\jkas{J. Korean Astron. Soc.}
\def\zpc{Z. Phys. C}
\def\phys{Physica}
\def\pr{Phys. Rep.}

\def\sal{Astron. Lett.}

\begin{document}
\title[]{Stability of magnetized strange quark matter in  the \\ MIT bag model with the density dependent bag pressure
}
\author{A. A.  \surname{Isayev}}
\email{alexanderi026@gmail.com}
  \affiliation{Kharkov
Institute of Physics and Technology, Academicheskaya Street 1,
 Kharkov, 61108, Ukraine }
 \affiliation{Kharkov National University, Svobody Sq., 4, Kharkov, 61022,
 Ukraine}
 \affiliation{Institute for the Early Universe,
 Ewha Womans University, Seoul 120-750, Korea}

\date{\today}

\begin{abstract}
The stability of magnetized strange quark matter (MSQM) is studied
in the MIT bag model with the density dependent bag pressure. In the
consistent thermodynamic description of  MSQM,  the quark chemical
potentials, the total thermodynamic potential  and the anisotropic
pressure acquire the corresponding additional term proportional to
the
density derivative of the bag pressure. 
The model parameter space is determined, for which MSQM is
absolutely stable, i.e., its energy per baryon is less than that of
the most stable $^{56}$Fe nucleus  under the zero external pressure
and vanishing temperature. It is shown that there exists the
magnetic field strength $H_{u\,max}$ at which the upper bound
$B_\infty^u$ on the asymptotic bag pressure $B_\infty\equiv
B(\varrho_B\gg \varrho_0$) ($\varrho_0$ being the nuclear saturation
density) from the absolute stability window vanishes. The value of
this field, \hbox{$H_{u\,max}\sim$$(1$--$3)\cdot10^{18}$}~G,
represents the upper bound
on the magnetic field strength, which can be reached in a strongly
magnetized strange quark star. It is clarified  how the absolute
stability window and upper bound on the magnetic field strength are
affected by varying the parameters in the Gaussian parametrization
for the density dependence of the bag pressure.
\end{abstract}



\maketitle

\section{Introduction and Basic Equations}
\label{I} After the conjecture that strange quark matter (SQM),
composed of deconfined $u$, $d$ and $s$ quarks, can be the ground
state of matter~\cite{AB,W,FJ}, it became the subject of intense
researches. In the astrophysical context, this would mean that the
formation of strange quark stars, made up entirely of SQM and
self-bound by strong interactions, is possible~\cite{I,AFO,HZS}. The
birth of a strange quark star can proceed via conversion of a
neutron star as a strong deflagration process during a few
milliseconds~\cite{PRD12Herzog}, accompanied by a powerful neutrino
signal~\cite{AL94Martemyanov}. If SQM is metastable at zero external
pressure, it can be encountered in the cores of heavy neutron stars
where the density of about several times nuclear saturation density
can be sufficient for the deconfinement phase  transition to
occur~\cite{FW}. Such stars, composed of the quark core and hadronic
crust, are called hybrid stars. Modern astrophysical observations,
including data on the masses and radii, spin-down rates, cooling
history, glitches and superbursts, do not disprove the existence of
quark matter in compact stars.

The other important feature is that compact stars can be endowed
with the strong magnetic field~\cite{DT}. Near the surface of
magnetars -- strongly magnetized  neutron stars -- the field
strength can reach values of about
$10^{14}$–-$10^{15}$~G~\cite{TD,IShS}. Even stronger magnetic fields
up to $10^{19}$-$10^{20}$~G can potentially occur in the cores of
neutron stars~\cite{CBP}. The large pulsar kick velocities due to
the asymmetric neutrino emission in direct Urca processes in the
dense core of a magnetized neutron star could be the possible
imprint of such ultrastrong magnetic
fields~\cite{Ch,ApJ95Vilenkin,HP,PRD99Arras}.
 The origin of magnetar's
strong magnetic fields is yet under discussion, and, among other
possibilities, it is not excluded that this can be due to
spontaneous ordering of nucleon~\cite{IY,PRC06I}, or quark~\cite{TT}
spins in the dense interior of a neutron star.

Strong magnetic fields can have significant impact on thermodynamic
properties of cold dense
matter~\cite{C,BCP,BPL,RPPP,IY4,IY10,WSYP,JPG14Dexheimer,arxiv14Chu}.
 In
particular, the pressure anisotropy, exhibited in the difference
between the  pressures along and perpendicular to the magnetic
field, becomes relevant for strongly magnetized
matter~\cite{Kh,FIKPS,IY_PRC11,IY_PLB12,NPA13SMS,JPG13IY}. In this
study, I consider  strongly magnetized SQM (MSQM) taking into
account the effects of the pressure anisotropy. 
I aim at finding the model parameter space for which MSQM is
absolutely stable, i.e., its energy per baryon is less than that of
the most stable nucleus $^{56}$Fe under the zero external pressure
and vanishing temperature. For the parameters from this absolute
stability window, the formation of a strongly magnetized strange
quark star is
possible. 

Note that in order to describe the confinement property of quantum
chromodynamics, in the conventional MIT bag model~\cite{CJJ} this is
achieved by introducing the density independent bag pressure by
which quarks
are confined in a finite region of space called a "bag". 
The standard thermodynamic equations can be used to study quarks
confined to a bag. Another phenomenological way to describe the
quark confinement is to consider the density dependent quark
masses~\cite{ZPhysC81Fowler,PLB89Chakrabarty,PRD95Benvenuto,PRC05Wen}.
In this case, an important issue of thermodynamic consistency
arises. Because of the density dependence of the quark masses, the
quark chemical potentials acquire an additional density dependent
term and become effective~\cite{PRC08Peng}. In fact, such
thermodynamically consistent approach was deve\-loped in the late
work~\cite{PRD14Xia}. Note also that the quark confinement was
modeled recently by the density- and isospin-dependent quark
masses~\cite{ApJ14Chu,PRD14Chu}.

These phenomenological QCD models were applied to study MSQM in
Refs.~\cite{C,BCP,RPPP,WSYP,JPG13IY,FMRO,HHRS,arxiv14Chu}. In
particular, the effects of the pressure anisotropy were disregarded
in Refs.~\cite{C,BCP,RPPP,WSYP}, in Refs.~\cite{FMRO,HHRS} only the
matter contribution to the pressure anisotropy was considered, and
both the matter and field contributions were accounted for in
Refs.~\cite{JPG13IY,PRD14Chu}. In this study, I consider the
absolute stability of MSQM in the MIT bag model with the density
dependent bag pressure $B(\varrho_B)$. The advisability to extend
the conventional MIT bag model came from the necessity to reconcile
the different constraints on the bag pressure at low and high baryon
densities obtained from heavy-ion experiments at CERN-SPS and
astrophysical observations of neutron stars with the masses well
above the mass of a canonical neutron star $M\sim 1.4M_\odot$
($M_\odot$ being the solar mass)~\cite{PLB02Burgio}. In other model
frameworks, the density dependent bag pressure was used in
Refs.~\cite{PhysRep93Adami,PLB99Blaschke}. Note that in the extended
MIT bag model, the quark chemical potentials  acquire the term
proportional to the density derivative $\frac{\partial
B}{\partial\varrho_B}$, and the issue of the consistent
thermodynamic description of MSQM becomes relevant. To that aim,  I
explore the formalism of the work~\cite{PRD14Xia}, developed
initially to describe the quark confinement by the density- (and/or
temperature-) dependent quark masses, and, after  proper
modification, apply it to the case of the extended MIT bag model.

Here MSQM will be considered as an uniform matter permeated by an
external uniform  magnetic field.  In the bag model, the matter part
of the total energy density (excluding the magnetic field energy
contribution) reads
\begin{align}E_m=\Omega_m^0+\sum_i \bar\mu_i\varrho_i \label{Em}, \end{align}
where
\begin{align}
\Omega_m^0=\sum_{i} \Omega_i^0+B(\varrho_B), \label{Om_m0}
\end{align}
and
$\Omega_i^0$ is the thermodynamic potential 
for free relativistic fermions of $i$th species ($i=u,d,s,e$) in the
external magnetic field, which is given by the same expression as in
Refs.~\cite{C,WSYP,JPG13IY} with the only difference that, according
to the approach of Ref.~\cite{PRD14Xia}, the real (nonrenormalized)
chemical potentials $\mu_i$ should be  substituted  there by the
effective (renormalized) chemical potentials $\bar\mu_i$.

At the given $H$, the
differential form of Eq.~\p{Em} is
\begin{align}\bigl(dE_m\bigr)_H=\bigl(d\Omega_m^0\bigr)_H+\sum_i
\bar\mu_i\,d\varrho_i
+\sum_i\varrho_i\,d\bar\mu_i\label{dEm},\end{align} where
$$\bigl(d\Omega_m^0\bigr)_H=\sum_i\frac{\partial\Omega_i^0}{\partial\bar\mu_i}\,d\bar\mu_i+\sum_i
\frac{\partial B}{\partial\varrho_i}\,d\varrho_i.$$ With account of
equation
\begin{equation}\varrho_i=-\Bigl(\frac{\partial\Omega_i^0}{\partial\bar\mu_i}\Bigr)_H\,, 
\label{ni0}
\end{equation}
Eq.~\p{dEm} acquires the form
\begin{align}\bigl(dE_m\bigr)_H=\sum_i \Bigl(\bar\mu_i +\frac{\partial B}{\partial\varrho_i}\Bigr)
\,d\varrho_i \label{dEmf}.\end{align} On the other hand, the
fundamental thermodynamic relation at zero temperature
reads~\cite{StPhysP1}
\begin{align}\label{fdEm}
    \bigl(dE_m\bigr)_H=\sum_i \mu_i d\varrho_i.
\end{align}
By comparing Eqs.~\p{dEmf} and \p{fdEm}, and taking into account
expression for the baryon number density
$\varrho_B=\frac{1}{3}(\varrho_u+\varrho_d+\varrho_s),
$ 
one gets the relationship between the real and effective chemical
potentials
\begin{align}
 \mu_e=\bar\mu_e,\quad   \mu_f=\bar\mu_f+\frac{1}{3}\frac{\partial
 B}{\partial\varrho_B},\;
 f=u,d,s.
 \label{mu_f}
\end{align}
Further I   study  charge neutral states of MSQM
 and
assume that the chemical equilibrium with respect to weak processes
is established among the fermion species with
the coresponding conditions on the real chemical potentials~\cite{AFO,PRD14Xia}. 
Note that, in view of Eqs.~\p{mu_f}, 
the effective
chemical potentials $\bar\mu_i$ satisfy the same equations, as the
real ones $\mu_i$:
\begin{align}\label{emud}
\bar\mu_d&=\bar\mu_u+\mu_{e^-},\quad 
\bar\mu_d=\bar\mu_s. 
\end{align}

 The Hugenholtz - van
Hove theorem establishes the thermodynamic relation between the
pressure and energy density at
zero temperature  for nonmagnetized fermion matter~\cite{Physica58Hugenholtz}. 
For magnetized fermion matter, the total pressure is the anisotropic
function of the magnetic field
strength~\cite{Kh,FIKPS,IY_PRC11,IY_PLB12,NPA13SMS,JPG13IY}. In
particular, the longitudinal $p^{\,l}$  and transverse $p^{\,t}$
pressures are different. By comparing expressions for the
longitudinal pressure $p^{\,l}$ 
and energy density~\cite{FIKPS,NPA13SMS,JPG13IY}, one can get the 
Hugenholtz - van Hove theorem  
for magnetized
matter in the form
\begin{equation}\label{prlm}
p^{\,l}_{m}=-E_m+\sum_i \mu_i\varrho_i,
\end{equation}
where $p^{\,l}_{m}$ is the matter part of the longitudinal pressure.
I will preserve this equation also for MSQM in the extended MIT bag
model. 
With
account of Eq.~\p{Em}, Eq.~\p{prlm} takes the form
\begin{equation}\label{plm}
    p^{\,l}_{m}=-\Omega_m^0+\sum_i
    (\mu_i-\bar\mu_i)\varrho_i=-\Omega_m^0+\varrho_B\frac{\partial B}{\partial\varrho_B}.
\end{equation}

At zero temperature, the matter part of the  thermodynamic
potential 
$\Omega_m$, determined according to the standard
thermodynamic equation
\begin{equation}
    \Omega_m=E_m-\sum_i\mu_i\varrho_i,
\end{equation}
with account of Eq.~\p{Em} becomes
\begin{equation}\label{Omm}
\Omega_m=\Omega_m^0-\sum_i
    (\mu_i-\bar\mu_i)\varrho_i=\Omega_m^0-\varrho_B\frac{\partial B}{\partial\varrho_B}.
\end{equation}
By comparing Eqs.~\p{plm} and \p{Omm}, one arrives at the
thermodynamic relationship 
$p^{\,l}_{m}=-\Omega_m.$
The matter parts of the longitudinal and transverse pressures are
related by the equation~\cite{NPA13SMS,PRD14Chu,PRD12SDM}
\begin{equation}\label{ptm}
p^{\,l}_{m}-p^{\,t}_{m}=HM,
\end{equation}
where $M=-\frac{\partial\Omega_m}{\partial H}$ is the system
magnetization. After summarizing Eqs.~\p{Em}, \p{Om_m0}, \p{plm},
\p{ptm}, and accounting for the pure magnetic field contribution,
the total energy density $E$, longitudinal $p^{\,l}$ and transverse
$p^{\,t}$ pressures for MSQM with  the density dependent bag
pressure can be written in the form
\begin{align}
E&=\sum_i \bigl(\Omega_i^0+\bar\mu_i\varrho_i\bigr)+\frac{H^2}{8\pi}+B,\label{E}\\
p^{\,l}&=-\sum_i\Omega_i^0-\frac{H^2}{8\pi}-B+\varrho_B\frac{\partial B}{\partial\varrho_B},\label{pl}\\
p^{\,t}&=-\sum_i\Omega_i^0-HM+\frac{H^2}{8\pi}-B+\varrho_B\frac{\partial
B}{\partial\varrho_B}. \label{pt}
\end{align}

In the case of the density independent bag pressure,
Eqs.~\p{E}--\p{pt} go over to the corresponding equations of
Ref.~\cite{JPG13IY}. Because of the breaking of the rotational
symmetry by the magnetic field, the longitudinal $p^{\,l}$ and
transverse $p^{\,t}$ pressures are not the same. There are two
different contributions to the pressure anisotropy: the matter
contribution proportional to the magnetization $M$, and the magnetic
field contribution given by the Maxwell term $\frac{H^2}{8\pi}$.
Note that it was argued recently~\cite{arxiv14Chatterjee} (and
discussed before in Refs.~\cite{JPC82Blandford,PRC12Potekhin}) that
the magnetization contribution to the energy--momentum tensor is
canceled by the Lorentz force associated with the magnetization
current, and, taking that into account, the pressure anisotropy
doesn't occur. In the given case, the direct answer on this question
is that I consider a spatially uniform distribution of the
  magnetic field and matter density, and, hence,
the magnetization current density
$\mathbf{j}_b=\nabla\times\mathbf{M}$ (in units with $c=1$) is
exactly zero. Therefore,  the associated Lorentz force density
$\mathbf{j}_b\times\mathbf{H}=0$,  and cannot compensate the term
with the magnetization in the transverse pressure. The other point
is that the main, most principal  source of the pressure anisotropy
is provided by the pure magnetic field contribution $H^2/8\pi$, and
there is no compensating effect for it. Just this term, as will be
shown later, 
plays the main role in establishing the upper bound
for the magnetic field strength in a 
magnetized medium. 

For a spatially nonuniform case, the argument, raised
in~\cite{arxiv14Chatterjee}, is based on the consideration of the
balance of the {\it volume} forces in the stationary state:
$\partial_l T^{kl}=0$,  $T^{kl}$ being the spatial components of the
quantum-statistical average
 of the energy-momentum tensor of the
system. Under writing the condition $\partial_l T^{kl}=0$ with
account of the Maxwell equation for the electromagnetic field
tensor, the Lorentz force density associated with the magnetization
current density, cancels the term with the magnetization, and the
corresponding first integral doesn't contain the magnetization as
well. While this step is correct,  it is then wrongly concluded in
Ref.~\cite{arxiv14Chatterjee}  that the system's equilibrium
(stationary state) is determined by the thermodynamic pressure and
any anisotropy in the matter pressure does not appear. Note that the
  pressures in the system are determined as the spatial
diagonal elements $T^{kk}$ of the energy-momentum tensor in the
system's rest frame. In the stable stationary state, they should be
positive $T^{kk}>0$. The conditions
 $T^{kk}>0$ and  $\partial_l T^{kl}=0$ are, obviously, different and {\it both} should be
satisfied in the stable stationary state.

In fact, the consistent derivation shows that the above mentioned
cancelation occurs, with account of the Maxwell equation, only in
the derivative $\partial_l T^{kl}$, but not in the energy-momentum
tensor itself. An example of the spatially uniform case,  discussed
above,  clearly confirms that.   The introduction by hand of the
Lorentz force contribution associated with the magnetization current
to the energy-momentum tensor is the artificial step, not confirmed
by the consistent derivation. Just the opposite, the consistent
microscopic derivations~\cite{FIKPS,PRD12SDM} show that such a
contribution is missing in the energy-momentum tensor, and, hence,
the anisotropy in the matter part of the total pressure is present
as well as the anisotropy caused by the contribution of the magnetic
field.

\section{Numerical results and discussion }
Now I will determine the absolute stability window of 
MSQM, subject to charge neutrality and chemical equilibrium
conditions, at zero temperature. The equilibrium conditions for
MSQM require vanishing 
the longitudinal $p^{\,l}$ and transverse $p^{\,t}$ pressures
\begin{align} \label{pll}
p^{\,l}=-\Omega=0, \quad
p^{\,t}=-\Omega+H\frac{\partial\Omega}{\partial
H}=0, 
\end{align} where $\Omega=\Omega_m+\frac{H^2}{8\pi}$ 
is the total thermodynamic potential 
of the system. In order to be absolutely stable, the energy per
baryon of MSQM
should be less than that 
of the most stable $^{56}$Fe nucleus under the equilibrium
conditions~\p{pll}. On the other hand, at $H=0$, the experimental
observation proves that two-flavor quark matter, consisting of $u$
and $d$ quarks, is less stable compared to $^{56}$Fe nucleus at zero
external pressure and vanishing temperature~\cite{FJ}. I will also
retain this constraint for strong magnetic fields
$H\gtrsim10^{17}$~G, although, strictly speaking, it is unknown
 from the experimental point of view
 whether two-flavor quark matter is less stable than
$^{56}$Fe nucleus under the equilibrium conditions \p{pll} 
in such strong fields. Thus, for determining the absolute stability
window of MSQM, I will use the following constraints:
\begin{equation}
\biggl.\frac{E_m}{\varrho_B}\biggr|_{uds}\leqslant\epsilon_H(^{56}\mathrm{Fe})
\leqslant
\biggl.\frac{E_m}{\varrho_B}\biggr|_{ud}\label{genst}.\end{equation}
Regarding, for the rough estimate, $^{56}$Fe nucleus as a system of
noninteracting nucleons, magnetic fields $H>10^{20}$~G are necessary
in order to significantly alter its energy per nucleon
$\epsilon_H(^{56}\mathrm{Fe})$ \cite{BPL}. Since I will consider
magnetic fields $H<5\cdot10^{18}$~G, further I use the approximation
$\epsilon_H(^{56}\mathrm{Fe})\approx
    \epsilon_0(^{56}\mathrm{Fe})=930\,\mathrm{MeV}$.

For determining the absolute stability window from
constraints~\p{pll} and 
\p{genst}, I utilize the Gaussian parametrization for the bag
pressure as a function of the  baryon density~\cite{PLB02Burgio}:
\begin{equation}\label{bagg}
    B(\varrho_B)=B_\infty+(B_0-B_\infty)\,
    e^{-\beta\bigl(\frac{\varrho_B}{\varrho_0}\bigr)^2},
\end{equation}
where $B_0=B(\varrho_B=0$),
$B_\infty=B(\varrho_B\gg\varrho_0$) %
($\varrho_0\simeq0.17$~fm$^{-3}$), and the parameter $\beta$
controls the rate of decrease of the bag pressure from $B_0$ to
$B_\infty$. In numerical analysis, I  use $\beta=0.01$ and
$\beta=0.17$.
Note that the equilibrium conditions~\p{pll} 
contain the derivatives $\frac{\partial B}{\partial H}$,
$\frac{\partial^2 B}{\partial\varrho_B\partial H}$. Further I will
assume that the magnetic field affects the bag pressures $B_0$ and
$B_\infty$ in the same way, and, hence, the difference $\Delta
B_0\equiv B_0- B_\infty$ is independent of $H$ and will be
considered as a parameter. Then the actual dependence of the bag
pressure $B$ on $H$ is contained in the asymptotic bag pressure
$B_\infty$: $\frac{\partial B}{\partial H}=\frac{\partial
B_\infty}{\partial H}$ while $\frac{\partial^2
B}{\partial\varrho_B\partial H}=0$. The absolute stability window
will be built in the plane $(H,B_\infty)$ under various fixed values
of 
$\Delta B_0$ and 
$\beta$. The upper
(lower) bound $B^u_\infty$ ($B^l_\infty$) on the asymptotic bag
pressure $B_\infty$ 
can be found from the first of the equilibrium conditions~\p{pll}:
\begin{align}\label{bui}
    B_\infty^{u\,(l)}(H)&=-\sum_{\begin{smallmatrix}
i=u,d,s,e\\ (i=u,d,e)
\end{smallmatrix} }
\,\Omega_i^0-\frac{H^2}{8\pi}\\ &\quad
    -\Delta B_0\,e^{-\beta\bigl(\frac{\varrho_B}{\varrho_0}\bigr)^2}
    \bigl(1+\frac{2\beta\varrho_B^2}{\varrho_0^2}\bigr)\nonumber
\end{align}
after  finding the 
effective chemical potentials $\bar\mu_i$ from the first constraint
to the left (right) in~\p{genst}, taken with the equality sign, and
charge neutrality 
and chemical equilibrium
conditions $\p{emud}$. 
 In Eq.~\p{bui}, the  baryon density $\varrho_B$ should be 
 determined from Eq.~\p{ni0} after finding the chemical potentials $\bar\mu_i$
at the given  $H$.

\begin{figure*}
\begin{center}
 \includegraphics[width=0.84\linewidth]{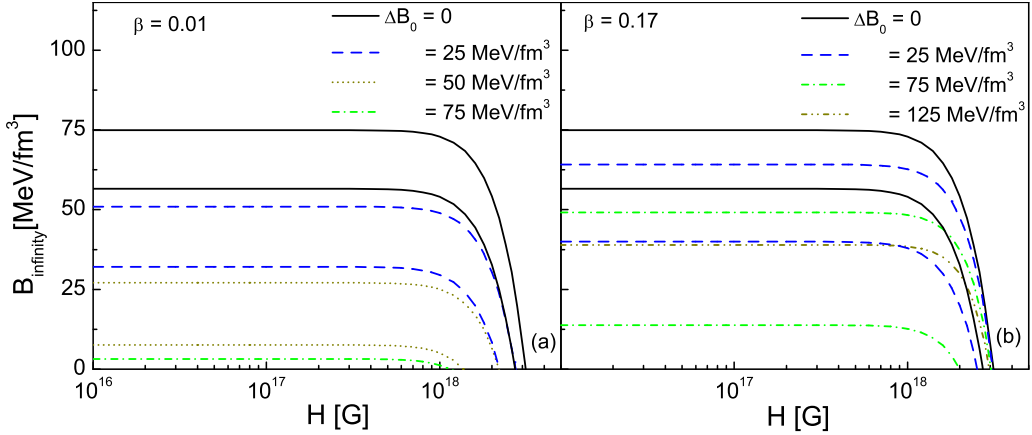}
\end{center}
\vspace{-4ex} \caption{(Color online) The absolute stability window
in the plane $(H,B_\infty$)  for MSQM at zero temperature with
$B(\varrho_B)$  given  by Eq.~\p{bagg} at (a) $\beta=0.01$ and (b)
$\beta=0.17$, and with a variable parameter $\Delta B_0$. The upper
$B_\infty^u$ and lower $B_\infty^l$ bounds are shown as the upper
and lower curves, respectively, in the pairs of the
similar curves 
(see also comments in the text).}
\vspace{-0ex}
\end{figure*}

Fig.~1 shows the dependences $B^u_\infty(H)$ and 
$B^l_\infty(H)$ for 
the
current quark masses $m_u=m_d=5$~MeV and $m_s=150$~MeV. 
 The upper
bound $B^u_\infty$ stays, at first, practically constant and then,
beginning from the magnetic field strength  $H$  somewhat smaller
than $10^{18}$~G,
 decreases.
 For example, the maximum
value of $B^u_\infty$, corresponding to $H=0$ (which is practically
indistinguishable from the value of $B^u_\infty$ at $H=10^{16}$~G)
is $B^{u}_{\infty,\,max}\approx 74.9$~MeV/fm$^3$ for $\Delta B_0=0$,
independently of the value of $\beta$; for $\Delta
B_0=75$~MeV/fm$^3$, $B^{u}_{\infty,\,max}\approx 3.2$~MeV/fm$^3$ at
$\beta=0.01$, and
 $B^{u}_{\infty,\,max}\approx
49.1$~MeV/fm$^3$ at $\beta=0.17$. The upper bound $B^u_\infty$
vanishes at $H_{u\,max}\approx3.1\times 10^{18}$~G for $\Delta
B_0=0$ at any $\beta$; for $\Delta B_0=75$~MeV/fm$^3$, $B^u_\infty$
vanishes at $H_{u\,max}\approx1.1\times 10^{18}$~G for $\beta=0.01$,
and at $H_{u\,max}\approx3\times 10^{18}$~G for $\beta=0.17$.  In
stronger magnetic fields $H>H_{u\,max}$, in order to satisfy the
equilibrium conditions, 
the
upper bound $B^u_\infty$ on the asymptotic bag pressure $B_\infty$
had to become negative, contrary to the constraint $B_\infty>0$.
 This means that, under the
equilibrium conditions and in magnetic fields $H>H_{u\,max}$, MSQM
cannot be absolutely stable.

The behavior of the lower bound $B^l_\infty(H)$ 
is similar to that of 
$B^u_\infty(H)$. 
At $H=0$, the maximum value of $B^l_\infty$ (which almost coincides
with the value of $B^l_\infty$ at $H=10^{16}$~G) is
$B^l_{\infty,\,max}\approx56.5$~MeV/fm$^3$ for $\Delta B_0=0$,
independently of the value of $\beta$; for $\Delta
B_0=25$~MeV/fm$^3$, $B^{l}_{\infty,\,max}\approx 32.0$~MeV/fm$^3$ at
$\beta=0.01$,  and $B^{l}_{\infty,\,max}\approx 40.0$~MeV/fm$^3$ at
$\beta=0.17$.   The lower bound $B^l_\infty$ stays practically
constant till magnetic fields somewhat smaller than $10^{18}$~G,
beyond which $B^l_\infty$ decreases. The lower bound $B^l_\infty$
vanishes at $H_{l\,0}\approx 2.7\times10^{18}$~G for $\Delta B_0=0$
at any $\beta$; for $\Delta B_0=25$~MeV/fm$^3$, $B^l_\infty$
vanishes at $H_{l\,0}\approx2.2\times 10^{18}$~G for $\beta=0.01$,
and at $H_{l\,0}\approx2.5\times 10^{18}$~G for $\beta=0.17$. Under
the equilibrium conditions and in the fields $H>H_{l\,0}$, the lower
bound $B^l_\infty$ would be negative. Because  $B_\infty>0$, the
inequality $B_\infty>B^l_\infty$ would be fulfilled always in the
fields $H>H_{l\,0}$.  Thus, in order for
 MSQM to be absolutely stable, the magnetic field strength
 should satisfy the constraint $H<H_{u\,max}$. In fact, the value $H_{u\,max}$
represents the upper bound on the magnetic field strength which can
be reached in a magnetized strange quark star. Note that the upper
bound $B^u_\infty$ decreases with increasing the parameter $\Delta
B_0$ at the given $\beta$. Hence, $B^u_\infty$ vanishes at a smaller
magnetic field for a larger value of $\Delta B_0$, i.e., the upper
bound on the magnetic field strength $H_{u\,max}$ decreases with
increasing $\Delta B_0$ at the given $\beta$. On the other hand, the
upper bound $B^u_\infty$ increases with increasing the parameter
$\beta$ at the given $\Delta B_0$. Hence, the larger 
$\beta$ is, the larger the upper bound 
$H_{u\,max}$  at the given $\Delta B_0$.

Note also that 
the  lower bound $B^l_\infty$
decreases with increasing the parameter $\Delta B_0$. Above certain
value of $\Delta B_0$, which depends on 
$\beta$, the
lower bound  $B^l_\infty$ would become negative,  and, hence, for
$B_\infty>0$ the inequality $B_\infty>B^l_\infty$ would be always
fulfilled at any $H$. 
For such $\Delta B_0$
and $\beta$, the absolute stability window corresponds to
$0<B_\infty<B^u_\infty$. This is just the case for $\Delta
B_0=75$~MeV/fm$^3$ at $\beta=0.01$, and for $\Delta
B_0=125$~MeV/fm$^3$ at $\beta=0.17$, shown in
Fig.~1a and Fig.~1b, 
respectively.
\begin{figure*}
\begin{center}
\includegraphics[width=0.84\linewidth]{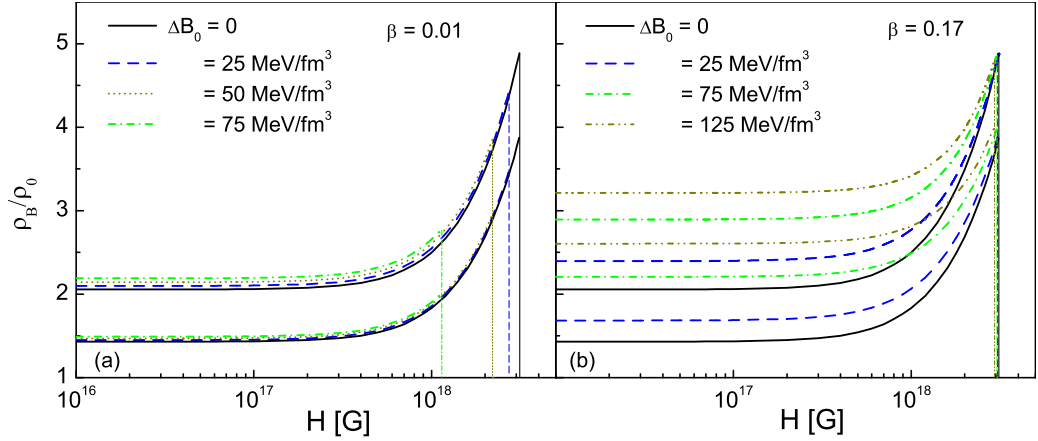}
\end{center}
\vspace{-2ex} \caption{(Color online)  The dependences 
$\varrho_B(H)$  
for MSQM
 and magnetized two-flavor quark matter (the upper  and lower curves, respectively, in the
pairs of the similar curves), determined under the 
equilibrium conditions at $E_m/\varrho_b=930$~MeV  for the same
values of the parameters $\Delta B_0$ and $\beta$ as in Fig.~1. The
bounding vertical lines on the right correspond to $H=H_{u\,max}$.}
\label{fig2}\vspace{0ex}
\end{figure*}

Fig.~2 shows the dependences $\varrho_b(H)$ for MSQM  and magnetized
two-flavor quark
 matter, determined under the respective equilibrium
conditions at $E_m/\varrho_b=930$~MeV, for magnetic fields
$H<H_{u\,max}$. In fact, the corresponding lines $\varrho_B^u(H)$
and $\varrho_B^{\,l}(H)$ represent the upper and lower bounds on the
baryon  density 
to ensure the absolute stability of MSQM.
 It is seen that the upper $\varrho_B^u$ and
lower $\varrho_B^{\,l}$ bounds  stay practically constant till the
magnetic field strength being somewhat smaller than~$10^{18}$~G, and
then increase till it reaches the corresponding maximum value. The
increase of the parameter $\Delta B_0$ at the given $\beta$, and the
increase of the parameter $\beta$ at the given $\Delta B_0$ lead to
the increase of
both the upper $\varrho_B^u$ and lower $\varrho_B^l$ bounds. 
For example, at $H=0$ the minimum value
of $\varrho_B^u$
 (being  almost the same as the value of $\varrho_B^u$ 
 at
 $H=10^{16}$~G) is
$\varrho_{B\,min}^u\approx2.1\varrho_0$ for $\Delta B_0=0$,
independently of $\beta$; for  $\Delta B_0=75$~MeV/fm$^3$,
$\varrho_{B\,min}^u\approx2.2\varrho_0$ at $\beta=0.01$, and
$\varrho_{B\,min}^u\approx2.9\varrho_0$ at $\beta=0.17$. Similarly,
at $H=0$ the minimum value of $\varrho_B^{\,l}$ is
$\varrho_{B\,min}^{\,l}\approx1.4\varrho_0$ for $\Delta B_0=0$,
independently of $\beta$; for  $\Delta B_0=75$~MeV/fm$^3$,
$\varrho_{B\,min}^{\,l}\approx1.5\varrho_0$ at  $\beta=0.01$, and
$\varrho_{B\,min}^{\,l}\approx2.2\varrho_0$ at  $\beta=0.17$.
 Note that magnetic fields $H\gtrsim10^{18}$~G
strongly affect the upper $\varrho_B^u$ and lower $\varrho_B^{\,l}$ bounds 
from the absolute stability window.

In conclusion, I have considered   MSQM under charge neutrality and
chemical equilibrium conditions in the MIT bag model with the
density dependent bag pressure $B(\varrho_B)$. 
  I aimed to determine the range for
the magnetic field strength $H$, asymptotic bag pressure
$B_\infty\equiv B(\varrho_B\gg \varrho_0)$, and  baryon density
$\varrho_B$, for which MSQM is absolutely stable, i.e., its energy
per baryon is less than that of the most stable $^{56}$Fe
nucleus  under the zero  pressure conditions \p{pll} 
and vanishing temperature.  In fact, this requirement sets the upper
bound on the parameters from the absolute stability window. The
lower bound  is determined from the constraint that magnetized
two-flavor quark matter under equilibrium conditions \p{pll} 
and zero temperature should be less stable than the most stable
$^{56}$Fe nucleus. This constraint is extended from the weak
terrestrial magnetic fields, where it has direct experimental
confirmation, to possible strong magnetar interior magnetic fields
$H\gtrsim10^{17}$~G, where such confirmation is wanting. It has been
shown that there exists the magnetic field strength $H_{u\,max}$ at
which the upper bound $B_\infty^u$ on the asymptotic bag pressure
$B_\infty$ from the absolute stability window vanishes. In fact, the
value of this field,
\hbox{$H_{u\,max}\sim$$(1$--$3)\cdot10^{18}$}~G, represents the
upper bound on the magnetic field strength, which can be reached in
a strongly magnetized strange quark star.  I have studied the effect
of the parameters in
the Gaussian parametrization 
 for the bag pressure on 
the absolute stability window and upper bound $H_{u\,max}$. 

It is interesting to note that the obtained  estimate for
$H_{u\,max}$ in strange quark stars
  is
similar  to the estimate $H\sim (1$--$3)\cdot10^{18}$~G for the
maximum average magnetic fields  in stable neutron stars, composed
of  strange baryonic matter~\cite{PLB02Broderick}. The found
estimate of the upper bound $H_{u\,max}$ 
in strange quark stars may
be further improved by including within 
the MIT bag model the effects of the perturbative quark
interactions~\cite{arxiv10Lattimer}. It would be of interest also to
extend this research to the case of the spatially nonuniform
magnetic field, whose realistic profile  should be determined from
the solution of the coupled Einstein and Maxwell
equations~\cite{ApJ15Belvedere}. The other interesting problem is to
take into account the vacuum corrections due to the magnetic field,
which are, however, beyond the scope of the MIT bag model. To that
aim, one can utilize   the quark chiral
models~\cite{PRD10Mizher,NPA03Ebert,PRC09Menezes}. Nevertheless, as
was shown in Ref.~\cite{PRC11Rabhi},  these corrections become
noticeable only in strong magnetic fields
$H\gtrsim3\times10^{19}$~G, and, therefore, one can expect that
their effect on the maximum magnetic field  in strange quark stars
will be of less importance.


\end{document}